# Unified understanding to the rich electronic-structure evolutions of 2D black phosphorus under pressure

Yu-Meng Gao, Yue-Jiao Zhang, Xiao-Lin Zhao, Xin-Yu Li, Shu-Hui Wang, Chen-Dong Jin, Hu Zhang, Ru-Qian Lian, Rui-Ning Wang, Peng-Lai Gong*, Jiang-Long Wang*, and Xing-Qiang Shi*

Key Laboratory of Optic-Electronic Information and Materials of Hebei Province, Hebei Research Center of the Basic Discipline for Computational Physics, College of Physics Science and Technology, Hebei University, Baoding 071002, P. R. China

*E-mails: gongpl@hbu.edu.cn, jlwang@hbu.edu.cn, shixq20hbu@hbu.edu.cn

**Abstract**: The electronic structure evolutions of few-layer black phosphorus (BP) under pressure shows a wealth of phenomena, such as the nonmonotonic change of direct gap at the Γ point, the layer-number dependence, and the distinct responses to normal and hydrostatic pressures. A full and unified understanding to these rich phenomena remains lacking. Here, we provide a unified understanding from the competition between *interlayer* quasi-bonding (QB) interactions and *intralayer* chemical bonding interactions. The former decreases while the latter increases the band gap under pressure and the origin can be correlated to different combinations of inter- and intra-layer antibonding or bonding interactions at the band edges. More interestingly, the interlayer QB interactions are a coexistence of two categories of interactions, namely, the coexistence of interactions between bands of the same occupancy (occupied-occupied and empty-empty interactions) and of different occupancies (occupied-empty interaction); and, the overall effect is a four-level interaction, which explains the anomalous interlayer-antibonding feature of the conduction band edge of bilayer BP. Our current study lay the foundation for the electronic structure tuning of two-dimensional (2D) BP, and, our analysis method for multi-energy-level interactions can be applied to other 2D semiconductor homo- and hetero-structures that have occupied-empty interlayer interactions.

# I. INTRODUCTION

Two-dimensional (2D) black phosphorus (BP) has attracted widespread attention for its unique properties and potential applications [1-5]. In monolayer BP, each phosphorus atom is covalently bonded to the three nearest neighboring atoms, and the periodic structure forms puckered hexagons. The monolayers are binding together to form a few-layer or bulk under the competition of interlayer van der Waals attraction (namely, the London dispersion force) and the quasi-chemical bonding (QB) repulsion [6]. BP shows significant in-plane anisotropy [7-9] and has direct gaps at the $\Gamma$ point ranging from 1.7 eV to 0.35 eV from monolayer to bulk [10, 11] with high carrier mobility [1]. These make BP promising for electronic and optoelectronic devices [12-16]. The band gap of few-layer BP can be modulated by external means such as doping [17-19], electric field [20-24], strain [10, 21, 25-28], and pressure [29-34]. Pressure changes the interlayer separations and intralayer bond lengths of phosphorus atoms, and the electronic structures are modified. The BP band structures have distinct responses to normal compressive strain and hydrostatic pressure [35, 36]. Normal strain leads to a direct-indirect band gap transition, and the indirect band gap close (semiconductor to metal transition) with increasing strain but the change in the direct gap is very small and may even increase a bit [35, 36]. Under hydrostatic pressure, a monotonic decrease of band gap is reported [29, 30].

At the interlayer region of 2D materials, in addition to the dispersion attraction between layers, there also exist interlayer QB interactions, namely, interlayer orbital-hybridization and the resulted energy-level splitting [37-40]. QB can cause energy-level splitting on the order of 1 eV [37]. Interlayer interactions can be classified into two main categories based on the occupancy of the involved energy bands close to Fermi energy: 1) interactions between bands of the same occupancy (occupied-occupied and empty-empty interactions) and 2) interactions between bands of different occupancies (occupied-empty interaction) [6]. The former *reduces* the band gap, as occupied-occupied interaction raises the valence band maximum (VBM) and empty-empty interaction lowers the conduction band minimum (CBM); while the latter *increases* the band gap, as occupied-empty interaction lowers VBM and raises CBM [6, 41]. In few-layer BP, these two categories of interlayer interactions may coexist due to the similar orbital character and the small energy-level separation between VB and CB, which is not identified to the best of our knowledge.

Under pressure, BP show complicated band gap evolution. For example, for the direct gap, experimental studies on BP have found that: the direct gap for monolayer BP increases monotonically under normal compressive strain while for bulk BP decreases monotonically under hydrostatic pressure, and for few-layer BP the direct gap exhibits nonmonotonic behavior (decreasing first and then increasing) [36]. In the so called "hydrostatic pressure" experiment in Ref. [36], the pressure on bulk BP is close to hydrostatic pressure; however, for monolayer and few-layer BP, due to the confinement of the substrate, the pressure that BP experience is more like a normal strain with fixed in-plane lattice constant. For the various electronic structure evolutions of BP under pressure in experiment (including under normal strain, the nonmonotonic variation of the direct gap and the direct-indirect band gap transition of few-layer BP and the monotonic decrease of the band gap under hydrostatic pressure), a unified understanding is still lacking.

In the current work, we develop a unified understanding to the different band gap variations in BP under normal and hydrostatic pressures through density-functional theory calculations and projected crystal orbital Hamilton population (*p*COHP) analysis [42]. We find that: 1) For the valence and conduction band edges evolution in bilayer (and few-layer) BP, the interlayer interactions are not simply a two-level interaction but involves four energy-levels because of the coexistence of the above-mentioned two main categories of interlayer-interaction. The four-level interaction can explain the abnormal interlayer-antibonding feature of the conduction band edge of the bilayer. 2) The overall effect of *interlayer* QB interactions lead to band gap reduction. In contrast, *intralayer* chemical bonding interactions lead to an increase in the band gap. Both of them are concluded by analyzing the inter- and intra-layer antibonding or bonding interactions at band edges. Under normal strain, the competition between them in few-layer BP results in a nonmonotonic variation of the direct gap; while for the indirect band gap evolution, the interlayer QB interactions dominate. 3) Under hydrostatic pressure, interlayer QB interactions lead to a monotonic decrease in the band gap of BP. 4) With increasing layer-number, the contribution of interlayer interactions increases and this results in the layer-dependent band gap evolution under strain. Our analysis method of multi-level interactions not only lay the foundation for tuning the electronic structure of few-layer BP but also can be applied to other 2D semiconductor homo- and hetero-structures involving occupied-empty interlayer interactions [6, 41]. The coexistence of two main categories of interlayer interactions could occur also for other systems with a similar orbital character in valence and conduction band edges.

## II. CALCULATION METHODS

Density-functional theory (DFT) [43] calculations were performed using the Vienna *ab initio* Simulation Package (VASP) [44, 45]. The projector augmented-wave (PAW) potentials were adopted to describe the core electrons [46, 47]. The valence electrons were described by plane-wave basis with an energy cut-off of 500 eV. The interlayer van der Waals interactions were included by the DFT-D3 method of Grimme *et al.* [48]. The 3D and 2D Brillouin zones were sampled by 10× 8 × 3 and 10× 8 $k$-point mesh (or $k$-point density of $2\pi \times 0.03$ Å$^{-1}$). For the simulations of few-layer, a vacuum of at least 15 Å along the z-axis was used to avoid interaction between periodic images of the slab model. The atomic positions were relaxed until the force on each atom was less than 0.02 eV/Å and the convergence criteria for energy were set to $10^{-5}$ eV. Electronic structure calculations adopted the hybrid functional of Heyd, Scuseria, and Ernzerhof (HSE06) [49]. Since the spin−orbital coupling (SOC) effect does not significantly change the electronic structure of BP, SOC was not included in calculation [50]. For bonding analysis of intralayer chemical-bonds and interlayer QB interactions, the LOBSTER package [42, 51] was used, which gives the crystal orbital Hamilton population (COHP) [52] via weighting the density of states (DOS) by the corresponding Hamiltonian matrix elements. For details of COHP calculations see Note S1 in the Supplemental Material [53]. The structures of monolayer and few-layer BP were extracted from the bulk BP under pressure [29]. This extraction method allows the obtained structure of monolayer and few-layer BP to include the pressure effect from other parts of the overall system, which approximates the effect of pressure from an inert pressure-transmitting medium, and the validity of this method has been demonstrated [29]. The simulation of normal strain was achieved by adjusting the interlayer spacing, the same to that did in literature [35]. The VASPKIT program was used for the post-processing of the electronic structure data [56].

## III. RESULTS AND DISCUSSION

Figures 1(a-b) shows the crystal structure of bulk BP, which also serves as the structural parents for few-layers under pressure. The structural parameters without pressure is $a$ = 3.31 Å, $b$ = 4.43 Å, and interlayer spacing $d$ = 3.19 Å, which agrees well with the optimized structure of bilayer BP [35]. Under a pressure larger than ~4.2 GPa, bulk BP undergoes a structural phase transition [57-60]. Here we only need to consider a pressure range of less than 2.5 GPa to probe the unified mechanism for various electronic-structure evolution. In this small range, one can estimate the normal strain by $P = (E - E_0)/[(L - L_0)A]$ [35], where $E$ and $E_0$ represent the energy of the system with and without applied strain, respectively, $L$ and $L_0$ the effective thickness of BP with and without applied strain, and $A$ the cell area in the lateral directions. In Fig. 1(a), $L$ represents the thickness of the bulk BP in one repeat unit; and for monolayer and few-layer ($N$ = 1, 2, 3), the effective thicknesses ($L_N$) is defined similar to that in bulk, namely, the maximum atomic-height-difference plus the interlayer spacing $d$. For bulk systems, the hydrostatic pressure can be calculated directly in DFT simulation.

### A. Analysis for nonmonotonic direct-gap evolution

BP has a direct gap at the Γ point. Under normal strain, the direct gap shows layer-dependent evolutions, as shown in Fig. 1(c). The calculated data points under normal strain are fitted using the same formula as in the literature [36], namely, use

$$\Delta E^N(P) = aP - \frac{\gamma_0}{2}\left(\sqrt{1 + \frac{P}{P_{\text{coh}}}} - 1\right)\cos\left(\frac{1}{N+1}\pi\right) \qquad (1)$$

for the fitting. Where $N$ represents the number of layers, $\Delta E^N$ represents the direct gap of $N$-layer BP; $a$ is the rate of change, $P$ is the pressure, $\gamma_0$ is the difference between the overlapping integral of the conduction band and the valence band at 0 GPa; $P_{\text{coh}}$ is known as cohesive pressure representing the threshold pressure that the BP layer needs to overcome during mechanical peeling. Using Eq. (1), the calculated results of different layers of BP were reproduced, as shown in Fig. 1 (c). The overall trend of the fitted curves is in good agreement with the calculated data and that in experiment [36].

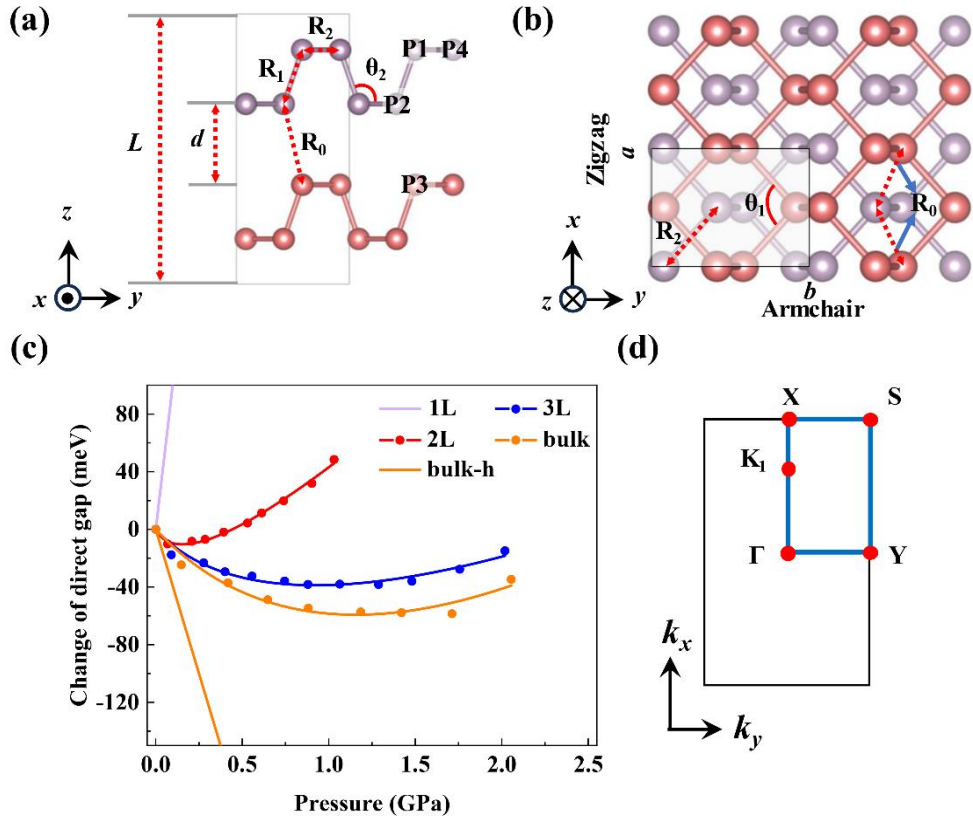

**FIG. 1. Structure and direct gap evolution under pressure.** (a) Side- and (b) top-views of bulk BP, which also serves as the structural parents for few-layers. In (a), $d$ and $L$ are the interlayer spacing and the cell length (or thickness of the bulk BP in one repeat unit) in the $z$ direction, respectively. In (a, b), the different P-P atom distances are labeled out, including the interlayer P2-P3 across the vdW gap ($R_0$), the intralayer P2-P1 along mainly the vertical direction ($R_1$), and the intralayer P1-P4 along the horizontal direction ($R_2$). $\theta_1$ and $\theta_2$ are the intralayer P-P-P bond angles. (c) Evolution of the direct gap under normal strain from monolayer to bulk. The line "bulk-h" means bulk BP under hydrostatic pressure. (d) The 2D Brillouin zone corresponding to the gray box in (b).

Under normal strain, we considered two possible conditions for the in-plane lattice constant: 1) fix the in-plane lattice constant as indicated in Ref. [36], and 2) optimize the in-plane lattice constant. The comparison of the two cases is shown in Fig. S1 in the Supplemental Material [53]. The variation of direct gap in Fig. S1 for monolayer and few-layer BP under the condition of fixed in-plane lattice constant is more consistent with the experimental results (Fig. 4 of Ref. [36]), and hence Fig. 1(c) displays the results from fixed in-plane lattice constant. For monolayer and few-layer BP, the pressure experiments close to the condition of fixed in-plane lattice constant (as discussed in the Introduction). From Fig. 1(c), it can be seen that the direct gap of the monolayer increases linearly under normal strain, while that of few-layer and bulk BP shows nonmonotonic variation, i.e., decreasing first and then increasing. More details, about the comparison between experiment and theory, can be found in Note S2 in the Supplemental Material [53].

In order to analyze the factors (interlayer and intralayer interactions) that govern the nonmonotonic change of direct gap under normal strain, we discuss in the following from the geometric and electronic aspects.

*Geometric aspect.* We analyze the structural change of BP from intralayer P-P bonds and interlayer P-P distances via the labeled $R_n$s ($n = 0$ to 2) in Figs. 1(a-b). $R_0$ denotes the interlayer P2-P3 nearest neighbor distance across the vdW gap, $R_1$ denotes the intralayer P2-P1 bond length mainly along the vertical direction and $R_2$ denotes the intralayer P1-P4 bond length. $\theta_1$ and $\theta_2$ are the intralayer P-P-P bond angles (will be used later).

We take bilayer BP as the example to probe the origin of the nonmonotonic variation of direct gap under normal strain. Table I summarizes the changes in interlayer P2-P3 distance ($R_0$), intralayer P2-P1 bond length ($R_1$ mainly along the vertical direction) and intralayer P1-P4 bond length ($R_2$) for bilayer BP under normal strain. As mentioned above, for the normal strain calculations, the in-plane lattice constant is fixed. The results reveal that the distance between atoms in the in-plane direction ($R_2$) remains essentially unchanged and the discussion can be simplified to the vertical direction. The $R_0$ and $R_1$ decreases in the out-of-plane direction with increasing normal strain. Table I shows that, under small strain (less than 0.21 GPa), only the interlayer $R_0$ decreases while the intralayer $R_1$ remains largely unchanged. Under normal strain of 1.03 GPa, the interlayer P2-P3 distance (intralayer P2-P1 bond length) decreases by 0.37 Å or 10% (0.02 Å or 1%).

**TABLE I.** Decreases in interlayer P2-P3 distance ($R_0$), intralayer P2-P1 bond length ($R_1$ mainly along the vertical direction) and intralayer P1-P4 bond length ($R_2$) of bilayer BP under normal strain. Refer to Figs. 1(a-b) for $R_0$, $R_1$, $R_2$ & P1, P2, P3, P4.

| P [GPa] | $R_0$ [Å] (P2-P3) | $R_1$ [Å] (P2-P1) | $R_2$ [Å] (P1-P4) |
|---|---|---|---|
| 0.00 | 3.67 | 2.26 | 2.22 |
| 0.07 | 3.62 | 2.26 | 2.22 |
| 0.21 | 3.58 | 2.26 | 2.22 |
| 0.29 | 3.54 | 2.25 | 2.22 |
| 0.39 | 3.50 | 2.25 | 2.22 |
| 0.53 | 3.46 | 2.25 | 2.22 |
| 0.61 | 3.42 | 2.25 | 2.22 |
| 0.74 | 3.38 | 2.24 | 2.22 |
| 0.90 | 3.34 | 2.24 | 2.22 |
| 1.03 | 3.30 | 2.24 | 2.22 |

In the following we show that, the nonmonotonic change of the direct gap of BP under normal strain can be understood from the combination of the changes in interlayer $R_0$ (intralayer $R_1$) and the corresponding bonding or antibonding characters in interlayer and intralayer along the vertical direction.

*Electronic aspect.* The VB and CB at the $\Gamma$ point (VB@$\Gamma$ and CB@$\Gamma$) of bilayer BP are mainly composed of out-of-plane $p_z$ orbitals and a small amount of in-plane ($s$, $p_y$) orbitals, as shown in Fig. 2(a) and Fig. S2 in the Supplemental Material [53]. Figure S2 also shows the corresponding band structure of monolayer BP as a reference for electronic structure analysis. From monolayer to bilayer, due to the interlayer QB interaction [37], the monolayer VB (CB) at $\Gamma$ point splits into VB and VB1 (CB and CB1) as labeled in Fig. 2(a). In the following, we focus on the band edges (VB@$\Gamma$ and CB@$\Gamma$) of bilayer BP.

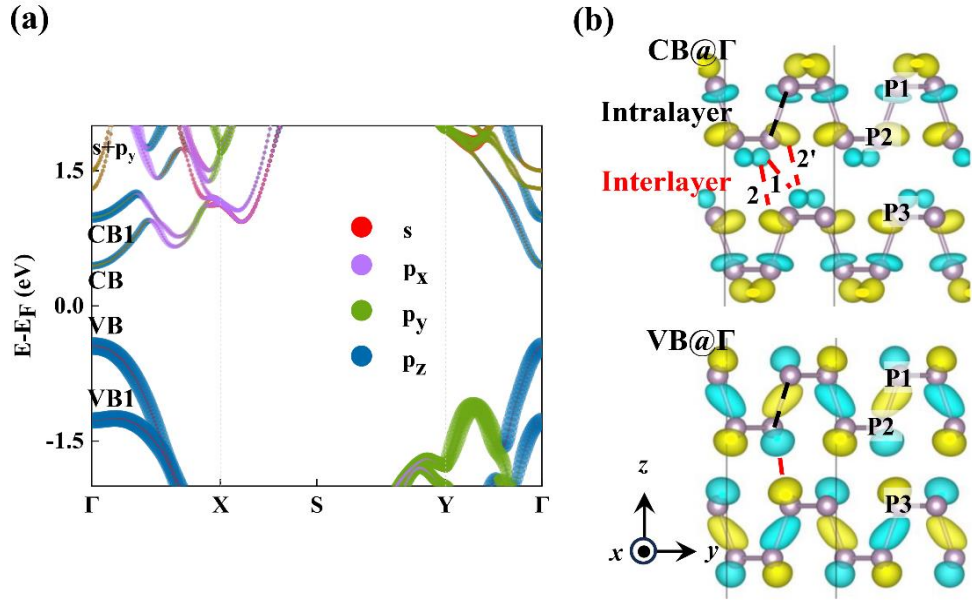

**FIG. 2. Inter- and intra-layer antibonding interactions of VB and CB edges for bilayer BP.** (a) Band structure projected to atomic-orbitals. The circle sizes are proportional to the weight of the projected orbitals. There is a color mixing for the $s + p_y$ orbirals as labeled at the upper-left of (a). (b) The CB and VB wave functions at $\Gamma$ point. The two colors of wave functions denote the opposite signs; red (black) dashed lines indicate the interlayer (intralayer) interactions between the P2-P3 (P2-P1) atoms; for the P2-P3 atoms across the vdW gap in CB@$\Gamma$, the labeled numbers (1, 2 and 2′) denote the coexistence of bonding (1) and antibonding (2 and 2′) interlayer interactions.

Figure 2(b) shows the real-space wave functions of VB@Γ and CB@Γ of bilayer BP (also refer to Figs. S3 and S4 for the monolayer BP wave functions as a reference in the Supplemental Material [53]). For the interlayer (P2-P3) QB interactions of bilayer BP, the VB@Γ exhibits an apparent interlayer-antibonding feature. However, for CB@Γ, the interlayer P2-P3 interactions involves three pairs of interactions as labeled by 1 (bonding) and 2 and 2′ (antibonding). Since bilayer BP [space group Pbcm (number 57)] has an inversion symmetry with the inversion center located at the midpoint of the line denoted by 1 in Fig. 2(b), the interlayer orbital-pair interactions of 2 and 2′ are equivalent and doubled with the inversion symmetry. The overall effect of interlayer interactions labeled by 1, 2, and 2′ is a weak antibonding (will be explained more in Table II). For the intralayer (P2-P1) chemical bonding interactions, the CB@Γ exhibits an apparent intralayer-antibonding feature. To determine the intralayer P2-P1 chemical bonding character for VB@Γ (and the above-mentioned interlayer P2-P3 QB character for CB@Γ), we adopt the COHP [52, 61] analysis by the LOBSTER package, in which the PAW wave functions of VASP are reconstructed to local-orbital basis [42, 51]. The COHP partitions the band-structure energy into the contributions of orbital-pair interactions, and the size of COHP indicate the strength of interlayer QB or intralayer chemical-bond interactions. More details see Note S1 in the Supplemental Material [53]. Table II lists the COHP for the interlayer (P2-P3) and intralayer (P2-P1) atom-pair interactions of CB@Γ, VB@Γ and CB@$K_1$. The COHP in Table II is a sum of the orbital-pair interactions in Tables SI, SII, and the Table included in Fig. S6 in the Supplemental Material [53]. The orbital-pair projected COHP (pCOHP) in Table SI show that, for the interlayer interaction of CB@Γ, the interactions between $p_z$–($s$, $p_y$) orbitals are also important apart from the $p_z$–$p_z$ interaction; and the same conclusion holds in Table SII for the intralayer interaction of VB@Γ. In Fig. 2(b), for the interlayer interaction of CB@Γ, the bonding interaction (labeled by 1) is the $p_z$–$p_z$ interaction, and antibonding interactions (labeled with 2 and 2′) are the $p_z$–($s$, $p_y$) interactions.

**TABLE II. The COHP analysis for the interlayer P2-P3 and intralayer P2-P1 interactions at Γ and $K_1$ points of bilayer BP.** A positive (negative) number means bonding (antibonding) and the size of numbers indicate the strength of interlayer QB (and intralayer chemical bonding) interactions. The numbers listed here is a sum of the orbital-pair interactions in Tables SI, SII, and the Table included in Fig. S6 in the Supplemental Material [53], At Γ point, the overall (summed) features of all interactions are antibonding.

| −COHP | Interlayer (P2-P3) | Intralayer (P2-P1) |
|---|---|---|
| CB@Γ | -0.023 | -0.615 |
| VB@Γ | -0.134 | -0.015 |
| CB@$K_1$ | 0.023 | -0.016 |

For the minus COHP (-COHP) numbers in Table II, a positive (negative) number means bonding (antibonding) and the modulus indicate the strength of interlayer QB (and intralayer chemical bond) interactions. Table II shows that the interlayer interaction of CB@Γ and intralayer interaction of VB@Γ are both (relative weak) antibonding. For monolayer BP, the intralayer P2-P1 interaction of VB@Γ is also weak antibonding (Fig. S4). The antibonding character of these interactions are not obvious from the wave function plots in Fig. 2(b) due to the multi-orbital character of each level (mainly $p_z$, $s$, and $p_y$ orbitals in Tables SI and SII), and hence the COHP analysis is a powerful tool to determine the bonding (antibonding) character and the relative strength of interactions for energy-levels with multiple atomic-orbitals.

For the CB@Γ of bilayer BP in Fig. 2(a), usually one thought it was a bonding level from the interlayer interaction from monolayer to bilayer (Fig. S2). However, the above COHP analysis show that the overall (summed) character is weak interlayer antibonding. The overall anti-bonding characteristic is anomalous, which indicate that the nature of interlayer QB interactions in BP is not simply the interaction between energy levels of the same occupancy (as sketched in Fig. 3a): namely, a multi-level orbital hybridization should be considered because of the coexistence of two main categories of interlayer-interactions owing to the similar orbital character and the small energy-level separation between VB and CB. Figure 3 indicates that, in addition to the two-level interaction between occupied-occupied and empty-empty interactions in Fig. 3(a), there is occupied-empty interaction as indicated in Fig. 3(b). The exact meaning of the occupied-empty interaction is shown in Fig. S5 in the Supplemental Material [53]. The overall effect is a four-level interaction in Fig. 3(c), which could lead to the weak antibonding character of CB in bilayer.

For the bilayer (2L) energy levels in Fig. 3, $VB_{2L}^{(2)}$ and $CB_{2L}^{(2)}$ denote the band edges from the (imagined) two-level interactions, and, $VB_{2L}^{(4)}$ and $CB_{2L}^{(4)}$ indicate that from the four-level interaction. In Fig. 3(a), the antibonding state of occupied-occupied interaction raises $VB_{2L}^{(2)}$ in bilayer relative to that in monolayer (1L) BP, and the bonding state of empty-empty interaction lowers $CB_{2L}^{(2)}$, making $VB_{2L}^{(2)}$ and $CB_{2L}^{(2)}$ close in energy. In addition, the VB and CB have similar orbital characters at Γ point [refer to Fig. 2(a)]. So, the occupied VB and empty CB can interact due to similar orbital character and close in energy. In Fig. 3(b), the interaction of VB and CB leads to the lowered $VB_{2L}^{(4)}$ and raised $CB_{2L}^{(4)}$. Namely, the additional occupied-empty interaction raises CB and lowers VB. So, the energy shift from the occupied-empty interaction in Fig. 3(b) is opposite to that in Fig. 3 (a); and, the final energy-shift in Fig. 3(c) is a combination of the effects from Figs. 3(a) and 3(b) -- that both VB and CB are raised in bilayer relative to that in monolayer. In addition, the four-level interaction also helps explain the CB wave-function evolution in real-space from 1L to 2L [Fig. 3(d)]: in 1L the two surfaces have the same real-space distributions of wave function while in 2L the inner surface and the outer surface is different due to the 1L VB mixes into 2L CB in the interlayer region. Till now, the four-level interaction picture explained the anomalous interlayer antibonding character of CB in bilayer.

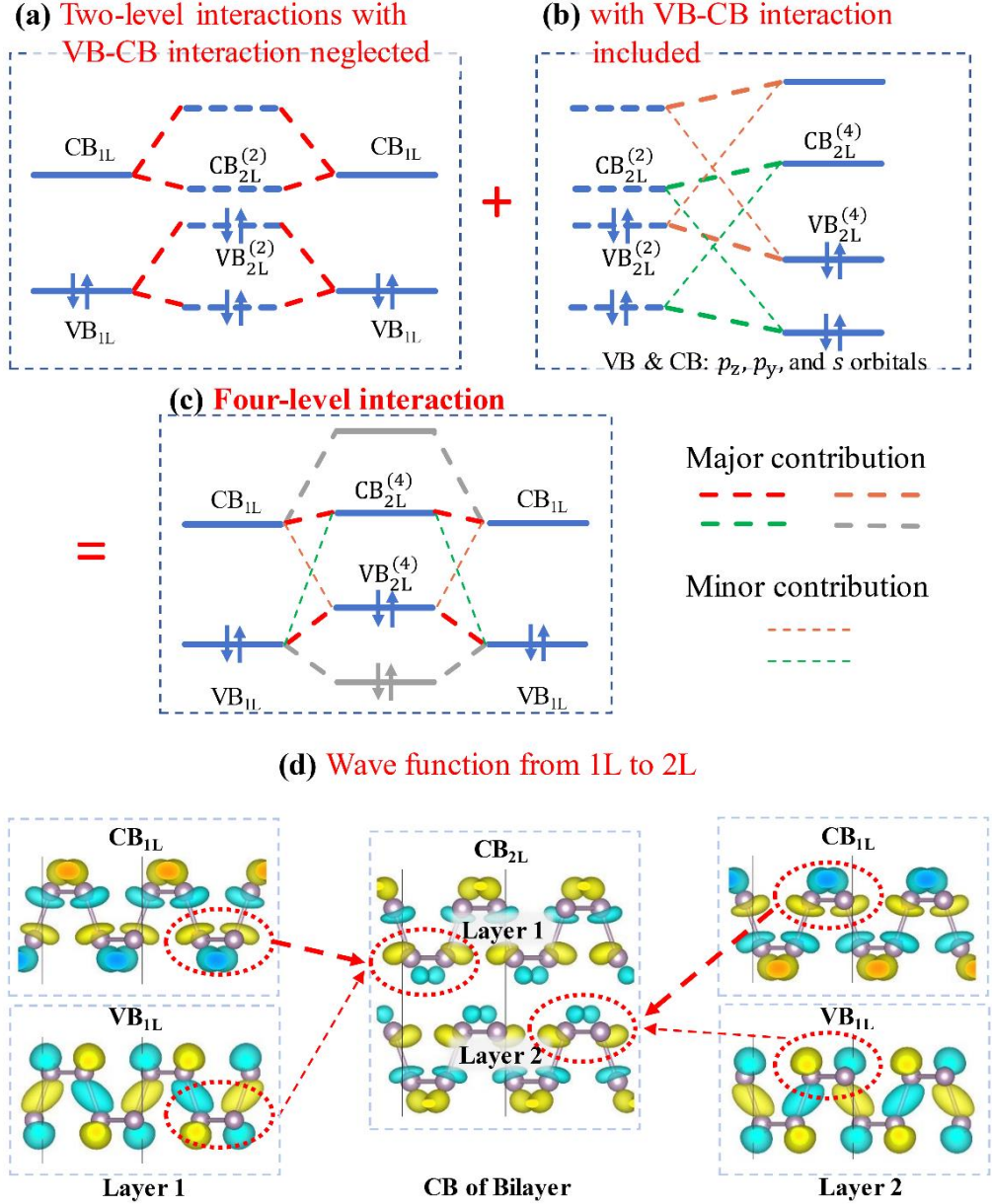

FIG. 3. Schematic diagram for the four-level interaction from monolayer (1L) to bilayer (2L) BP. In addition to the two-level interactions of occupied-occupied and empty-empty levels in (a), there is occupied-empty interaction [62] in (b) since they have similar orbital characters ($p_z$, $p_y$ and $s$ orbitals); and the overall effect is the four-level interaction in (c). For the bilayer energy levels, $VB_{2L}^{(2)}$ and $CB_{2L}^{(2)}$ denote the band edges from the (imagined) two-level interactions, and, $VB_{2L}^{(4)}$ and $CB_{2L}^{(4)}$ indicate the final band edges from the four-level interaction. Panel (c) is a simplified description of the four-level interaction, namely, at least the indicated interactions are needed for understanding the antibonding feature of $CB_{2L}^{(4)}$. (d) The 2L CB wave function in real-space with major contribution from 1L CB wave function (for outer part and interlayer region of 2L) and minor contribution from 1L VB (mainly for the interlayer region of 2L).

*Effects of interlayer and intralayer interactions.* From Tables I and II, one can analyze the effect of inter- and intra-layer interactions on band-edge (VB@Γ, CB@Γ) evolutions under strain [Fig. 4(b)], and then understand the direct gap evolution under strain [Figs. 4(a) and 1(c)]. Under strain, the interlayer $R_0$ and intralayer $R_1$ decrease (Table I), indicating that the inter- and intra-layer antibonding interactions (of P2-P3 and P2-P1) are enhanced. Table II shows that, for bilayer BP the inter- and intra-layer interactions of VB@Γ and CB@Γ all exhibit (overall) antibonding features but with different strength. For the intralayer (P2-P1) interactions, CB@Γ has significant antibonding while VB@Γ shows weak antibonding (Table II), which indicate the effect of intralayer interactions under strain is to increase the direct gap since the CB@Γ energy up-shift is larger than VB@Γ, as indicated in Fig. 4(b). Similarly, the effect of interlayer interactions under strain is to decrease the direct gap due to the energy up-shift of VB@Γ is larger than CB@Γ since the VB@Γ has significant antibonding while CB@Γ shows weak antibonding [Table II and Fig. 4(b)]. Under normal strain, both interlayer $R_0$ and intralayer $R_1$ have changed. In our calculations, it is possible to modify only $R_0$ ($R_1$) while keeping $R_1$ ($R_0$) fixed, thereby allowing for separately analyzing the influence of interlayer (intralayer) interactions on band evolution under strain. In Fig. 4(a), the variation of the direct gap (the blue line) is decomposed into the effects of interlayer (P2-P3) QB interactions (the red line) and intralayer (P2-P1) chemical bonding interactions (the black line). As discussed above for Table I, under small strain only the interlayer $R_0$ decreases while the intralayer $R_1$ remain unchanged. This results in the unchanged direct gap under small strain for the intralayer contribution [the black line in Fig. 4(a)]. As the strain increases, both the interlayer P2-P3 distance $R_0$ and the intralayer P2-P1 bond length $R_1$ decrease, resulting in the decrease in the direct gap from the interlayer interactions [red line in Fig. 4(a)] and the increase in the direct gap from the intralayer interactions [black line in Fig. 4(a)]. The overall effect [the blue line in Fig. 4(a)] is a nonmonotonic variation of the direct gap, namely, decreases first and then increases.

Similar phenomena of the nonmonotonic direct gap variation are observed in trilayer and bulk BP, as shown in Figs. 4(c, d). For monolayer BP, since there is only intralayer interactions, the direct gap increases monotonically [Fig. 1(c)]. Under normal strain, the change of the direct gap of BP exhibits a pronounced layer-number-dependence [Figs. 4(a, c, d) and 1(c)]. The layer-dependent effects can be attributed to the changes in the proportions of interlayer and intralayer contributions, namely, from bilayer to trilayer to bulk, the ratio of layer-number (intralayer interaction) and number of vdW gaps (interlayer interaction) changes from 2:1 to 3:2 to 1:1.

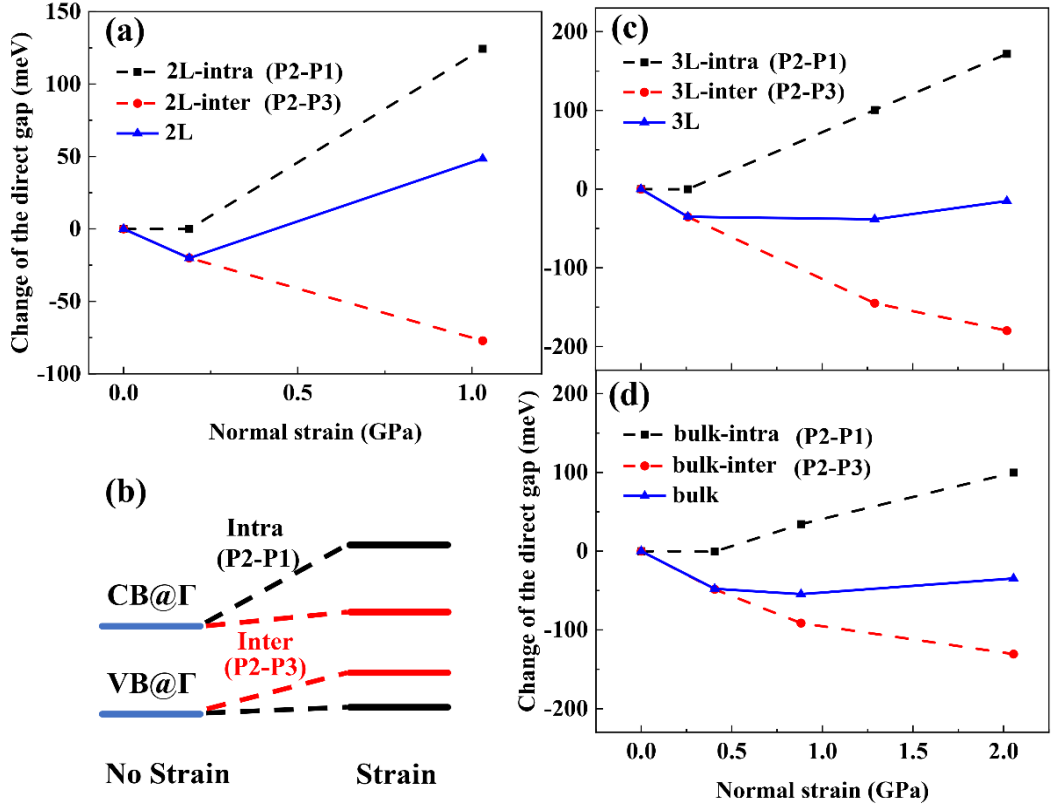

**FIG. 4. Variation of the direct gap for BP under normal strain.** For (a) bilayer, (c) trilayer, and (d) bulk BP, the overall gap change (the blue line) is decomposed into the effects of interlayer P2-P3 QB interactions (the red line) and intralayer P2-P1 chemical bonding interactions (the black line), both of them are from auxiliary calculations (see text for details); (b) the effects of inter- and intra-layer interactions under strain on the evolution of band edges, which determines the variation of the direct gap in (a), and also (c, d).

### B. Apply to direct-indirect gap transition

Under normal strain, in addition to the nonmonotonic change of the direct gap at the $\Gamma$ point, a direct-to-indirect band gap transition occurs in the entire Brillouin zone of few-layer BP, as shown in Table III and Figs. 5(a-b) using bilayer BP as an example. This is consistent with the literature reported results [35]. Here we demonstrate that our above analysis method can be applied also to this.

**TABLE III. Changes in the direct and indirect band gap sizes ($\Delta E$) under normal strain (from 0 GPa to 1.03 GPa) for bilayer BP.**

| 2L BP | 0 GPa | 1.03 GPa | $\Delta E$ |
|---|---|---|---|
| Direct gap | 0.891 eV | 0.940 eV | +49 meV |
| Indirect gap | 1.107 eV | 0.556 eV | −551 meV |

Under normal strain, apart from the meV-level variations in the direct gap [Fig. 1(c) and Table III], there is a more significant change in the CBM at the $K_1$ point (denoted as CB@$K_1$), refer to Figs. 5(a, b).

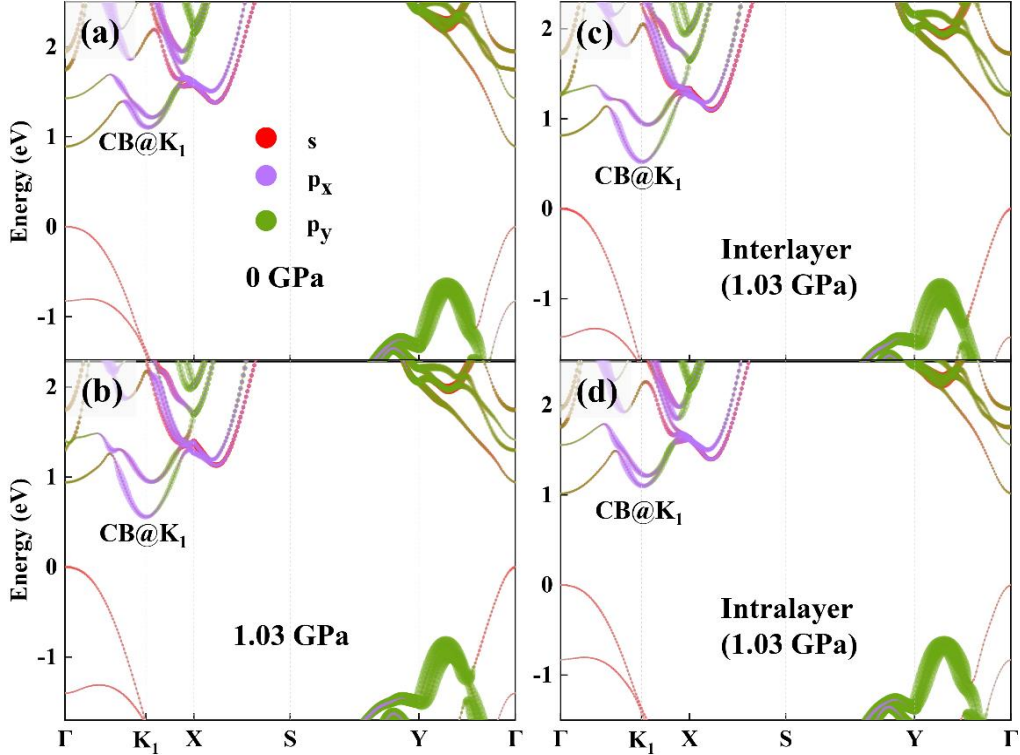

**FIG. 5. Direct-indirect band gap transition of CBM from $\Gamma$ point to $K_1$ point under normal strain.** (a) and (b) show the band-structure change of bilayer BP from 0 GPa to 1.03 GPa. (c, d) Decompose the band structure change to interlayer QB interactions (c) and intralayer chemical bonding interactions (d) at 1.03 GPa.

Figs. 5(a, b) show that CB@$K_1$ decreases significantly under strain and becomes lower than CB@$\Gamma$ in Fig. 5(b). Based on the analysis method in the above Subsection, we know that under normal strain, the decreases in interlayer $R_0$ and intralayer $R_1$ influence CB@$K_1$. Through QB and bonding analysis for CB@$K_1$ (Table II and Fig. S6 in the Supplemental Material [53]), we find that CB@$K_1$ shows interlayer (P2-P3) bonding features and intralayer (P2-P1) antibonding features, and the latter is relatively weak. In Figs. 5(c, d), by decomposing the structural change to the interlayer QB interactions and intralayer chemical bonding interactions (as does before in Fig. 4): it can be seen that the interlayer QB interactions mainly affect CB@$K_1$. As shown in Table II, Fig. S6 and Fig. S7 in the

Supplemental Material [53], the CB@$K_1$ exhibits interlayer (P2-P3) bonding character. The main contribution at the $K_1$ point comes from in-plane $p$ orbitals and $s$ orbital, as shown in Fig. S2. The change in CB@$K_1$ can be understood as the result of interlayer bonding interaction formed by the two-level (empty-empty) interaction of in-plane orbitals, similar to that shown in Fig. 3(a). Under normal strain, due to the interlayer bonding interactions, the CB@$K_1$ is lowered and the indirect band gap decreases in Table III.

In summary, the interlayer distance $R_0$ and intralayer bond length $R_1$ decreases under normal compressive strain and hence the inter- and intra-layer antibonding or bonding interactions (of P2-P3 and P2-P1) are enhanced. At the $K_1$ point, the traditional two-level interaction lowers the CB edge at $K_1$ under normal strain [63]. At the Γ point, however, due to the competition between inter- and intra-layer interactions, the direct gap shows a nonmonotonic evolution and the change in the direct gap size is small for bilayer and trilayer BP [Table III and Fig. 1(c)]. More interestingly, at the Γ point the interlayer interaction is a four-level interaction [Fig. 3(c)], which explains the abnormal interlayer (P2-P3) antibonding character of CB@Γ [Table II and Fig. 2(b)]. In the following, we show that our analysis method also applies to the monotonic gap decrease under hydrostatic pressure.

### C. Apply to monotonic gap decrease

Above is about normal strain, now we move to hydrostatic pressure. For bulk BP (thickness of ~1 $\mu$m in the pressure experiment [36]), the pressure on the side surfaces can be large enough to loosen the contact with the diamond surface, resulting in a true hydrostatic pressure effect [36]. As the hydrostatic pressure increases, the band gap of bulk BP decreases monotonically, as shown in the above Fig. 1(c) with the line labeled by "bulk-h".

Under hydrostatic pressure, BP maintains its direct gap at the Γ point (before gap closing), and the band gap decreases monotonically [29, 30]. Under hydrostatic pressure, both few-layer and bulk BP exhibit a monotonically decreasing band gap, as shown in Fig. 6(a), and is consistent with previous theoretical calculations [29, 30]. The analysis method for the pressure effect on band gap evolution with hydrostatic pressure can be the same as in the above case of normal strain, namely, study the effect of inter- and intra-layer interactions on VB@Γ and CB@Γ evolutions under pressure.

Here, we also take bilayer BP as an example to investigate the origin of the monotonic reduction of band gap under hydrostatic pressure. Table IV summarizes the crystal structure changes in interlayer and intralayer for bilayer BP under hydrostatic pressure. In the normal direction across the vdW gap, the interlayer P2-P3 distance $R_0$ decreases (by 0.21 Å or 6%) under hydrostatic pressure of 2 GPa (Table IV), while the intralayer $R_1$ has remained essentially unchanged. The bond angle $\theta_2$ [refer to Fig. 1(a)] is decreased by 1.85° or 2% under hydrostatic pressure of 2 GPa, which leads to significant changes in lattice constant $b$ in Table IV. The lattice constant $a$ has remained essentially unchanged. So, for structural changes that related to band gap decrease under hydrostatic pressure, we only need to focus on the interlayer $R_0$ (P2-P3) in Table IV, since the bond angle change is less important for electronic structure change than bond length change.

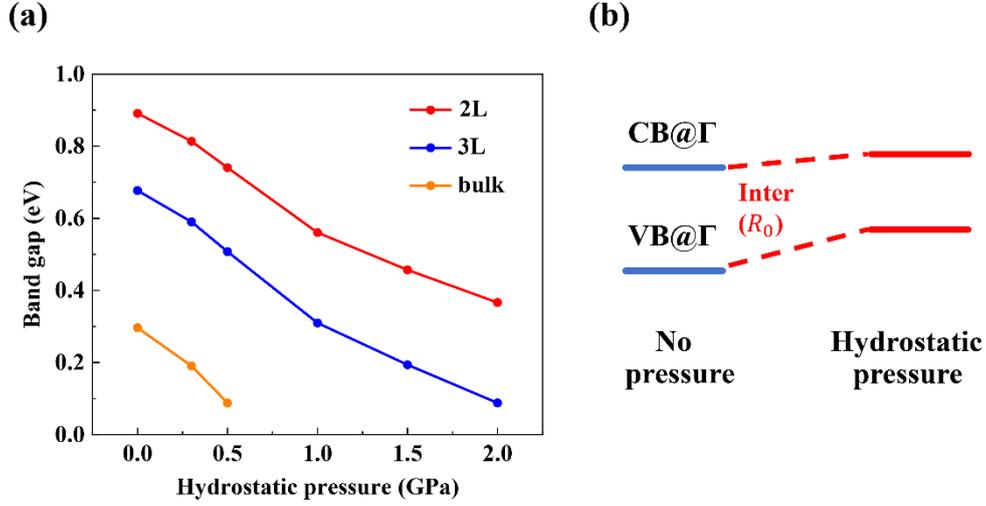

**FIG. 6. Band gap change under hydrostatic pressure.** (a) Evolution of the band gap under hydrostatic pressure from few-layer to bulk. (b) The effects of interlayer interactions under pressure on the evolution of band edges, which finally determines the evolution of band gap in (a).

Under pressure the interlayer interactions enhanced due to interlayer $R_0$ decrease (Table IV). From the previous analysis in the Subsection A for interlayer $R_0$ (P2-P3 interaction), we know that, the effect of interlayer interactions under pressure is to decrease the band gap due to the energy up-shift of VB@Γ is larger than CB@Γ since the VB@Γ has significant antibonding while CB@Γ shows weak antibonding [Table II and Fig. 6(b)]. As hydrostatic pressure increases, the interlayer P2-P3 distance $R_0$ decrease, resulting in the monotonic decrease in the band gap of few-layer and bulk BP from the interlayer interaction, as shown in Fig. 6(a).

**TABLE IV.** Changes in interlayer P2-P3 distance ($R_0$), intralayer P2-P1 bond length ($R_1$ mainly along the vertical direction), and other structural parameters of bilayer BP under hydrostatic pressure. Refer to Figs. 1(a-b) for the meaning of the structural parameters. The meaning of $c$ is equivalent to the effective thickness of bilayer BP.

| P[GPa] | $R_0$ [Å] (P2-P3) | $R_1$ [Å] (P2-P1) | $\theta_1$ [°]/$\theta_2$ [°] | $a$ [Å] | $b$ [Å] | $c$ [Å] |
|---|---|---|---|---|---|---|
| 0.00 | 3.67 | 2.26 | 96.22/102.46 | 3.31 | 4.43 | 10.65 |
| 0.30 | 3.65 | 2.26 | 96.11/102.15 | 3.30 | 4.39 | 10.63 |
| 0.50 | 3.61 | 2.26 | 96.30/101.84 | 3.31 | 4.35 | 10.57 |
| 1.00 | 3.55 | 2.26 | 96.33/101.28 | 3.31 | 4.28 | 10.45 |
| 1.50 | 3.50 | 2.26 | 96.39/100.92 | 3.31 | 4.24 | 10.36 |
| 2.00 | 3.46 | 2.25 | 96.45/100.61 | 3.31 | 4.20 | 10.27 |

## IV. CONCLUSION

In summary, we have provided a unified understanding of the various changes in the band structure of BP under pressure based on the changes in interlayer $R_0$ (intralayer $R_1$) and the corresponding bonding or antibonding characteristics in interlayer and intralayer along the vertical direction. We have elucidated the effects of interlayer QB interactions and intralayer chemical bonding interactions on the band gap of few-layer BP under pressure. The former decreases the band gap, while the latter increases the band gap under pressure. For the interlayer QB interactions, two main categories of interactions coexist (namely, interactions between bands of the same occupancy and bands of different occupancies coexist), the overall effect is a four-level interaction, which explains the abnormal interlayer-antibonding feature of CB@$\Gamma$ in the bilayer. Under normal strain and hydrostatic pressure, the direct gap dominated by out-of-plane $p_z$ orbitals and a small amount of in-plane ($s$, $p_y$) orbitals, while the CB@$K_1$ dominated by in-plane orbitals under normal strain. Under normal strain, the competition between interlayer QB interactions and intralayer chemical bonding interactions leads to a nonmonotonic variation in the direct gap at the $\Gamma$ point of few-layer BP, while for the indirect band gap (related to CB@$K_1$) the interlayer QB interactions dominate and causes the decrease of indirect band gap. The variation of BP band gap under hydrostatic pressure is dominated by a single factor -- the interlayer interaction -- which leads to the decrease of band gap, and hence BP shows a monotonic decrease in the band gap under hydrostatic pressure. Under strain, the change of the direct gap of BP exhibits a pronounced layer-dependent effect due to the increase in the proportion of interlayer contribution. A multi-level and multi-orbital analysis method is developed, which provides a unified explanation for the various electronic-structure evolutions under pressure. Moreover, this analysis method of interlayer-multilevel-interaction can also be applied to other 2D layered materials that has interlayer-interactions between occupied and empty states [64], and the coexistence of different categories of interlayer interactions that could occur also for other systems with a similar orbital character in valence and conduction band edges.

## ACKNOWLEDGMENTS

This work was supported by the National Natural Science Foundation of China (Grants Nos.

12274111 and 12104124), the Central Guidance on Local Science and Technology Development Fund Project of Hebei Province (236Z0601G), the Natural Science Foundation of Hebei Province of China (Grants Nos. A2021201001 and A2021201008), the Scientific Research and Innovation Team of Hebei University (Grant No. IT2023B03), the Advanced Talents Incubation Program of the Hebei University (Grants Nos. 521000981390, 521000981394, 521000981395, 521000981423, and 521100221055), and the high-performance computing center of Hebei University.

Y. -M. Gao and Y. -J. Zhang contributed equally to this work.

*Supplemental Material of:*

# Unified understanding to the rich electronic-structure evolutions of 2D black phosphorus under pressure

Yu-Meng Gao, Yue-Jiao Zhang, Xiao-Lin Zhao, Xin-Yu Li, Shu-Hui Wang, Chen-Dong Jin, Hu Zhang, Ru-Qian Lian, Rui-Ning Wang, Peng-Lai Gong*, Jiang-Long Wang*, and Xing-Qiang Shi*

Key Laboratory of Optic-Electronic Information and Materials of Hebei Province, Hebei Research Center of the Basic Discipline for Computational Physics, College of Physics Science and Technology, Hebei University, Baoding 071002, P. R. China

*E-mails: gongpl@hbu.edu.cn, jlwang@hbu.edu.cn, shixq20hbu@hbu.edu.cn

## Contents (Note, Tables, and Figures):

**Note S1. More on COHP analysis:**

For bonding analysis of inter- and intra-layer interactions, the LOBSTER package [1, 2] was used, which gives the crystal orbital Hamilton population (COHP) [3] via weighting the density of states (DOS) by the corresponding Hamiltonian matrix elements. The LOBSTER package is based on an analytic projection from projector-augmented wave (PAW) density-functional theory computations, reconstructing chemical information in terms of local, auxiliary atomic orbitals with the requirement that the PAW band functions coincide with the projected band functions, and then calculating the projected-COHP (*p*COHP) matrix elements based on the atomic orbitals[2, 4, 5]. It uses improved (augmented) Slater-type orbitals and exploits an orthonormalization technique [1], and the projection quality is measured by the absolute charge spilling parameter for occupied states and the absolute total spilling for all levels including unoccupied states, which are averaged over bands and *k*-points [1]. In our calculation, the absolute charge spilling was less than 2.48%. The orbital-pair projected COHP (*p*COHP) is defined as [2]:

$$-\text{COHP}_{ij}(E) = H_{ij} \sum_n \Re(c_i^n c_j^{*n}) \delta(E - E_n)$$

where $c_i^n$ are the coefficients associated with the atomic orbitals $\phi_i$ in a molecular orbital $\psi_n = \sum_i c_i^n \phi_i$, $H_{ij}$ is the Hamiltonian matrix element between the atomic orbitals $\phi_i$ and $\phi_j$, which are the *s*- and three *p*-orbitals (for valence electrons) of phosphorus atoms in our calculation, $\Re$ means the real part, $E_n$ is the energy levels of CB, VB at Γ point as well as CB at the $K_1$ point respectively as marked in Fig. 2(a) and Fig. 5(a) of the main text. The *p*COHP divides band structure energy into orbital-pair interactions. The *p*COHP summed over i, j for a given atom-pair provides indications for the strength of the bonding/anticonding interactions between the two atoms. The bonding is characterized by a positive overlapping population, and the off-site element (between different atoms) in the corresponding Hamiltonian will be negative. To make the *p*COHP numbers easier to understand, the data used are minus *p*COHP (-*p*COHP) so that positive means bonding, negative means antibonding, and zero means non-bonding (no interaction). For the analysis of bands near Fermi energy for a certain *k*-point, the band- and *k*-resolved *p*COHP was calculated [5, 6].

**Note S2. The discussion of comparing theory and experiment:**

Comparing theory and experiment, there may be two main aspects of differences.

1) In Ref. 36, for thicker BP, such as bulk BP (thickness of ~1 μm), the pressure on the side surfaces can be large enough to loosen the contact with the diamond surface, resulting in a true hydrostatic pressure effect. As the hydrostatic pressure increases, the band gap of bulk BP decreases monotonically, as shown in the above Fig. 1(c) in our manuscript with the line labeled by "bulk-h". From monolayer to bulk, as the number of layers increases, the effect of hydrostatic pressure gradually becomes more apparent, that is, the effect of hydrostatic pressure depends on the number of layers and for a few-layer more like normal strain but with mixing of hydrostatic pressure. The mixing of the effect of hydrostatic pressure could make the curve less steeper and close to the bulk results.

2) In the experiment of Ref. 36, few-layer BP were peeled off from the bulk and transferred to the surface of the diamond anvil cell (DAC). To avoid degradation, silicone oil was chosen as an inert pressure transmitting medium (PTM). For 2L BP, one outer surface is in contact with the diamond surface and the other outer surface is in contact with silicone oil. However, in terms of calculation, both outer surfaces are directly in vacuum, which may be the reason why the calculation is significantly different from the experiment (much steeper) for 2L BP. For 3L BP, however, due to the presence of the inner layer BP contact with the upper and lower BP layers, and hence the inner BP layer has the same environment as that in the experiment. This may be the reason why 3L is more consistent with the experiment than 2L.

In summary, due to the small energy scale (~ 0.05 eV) in discussion, all the above details may affect the small energy changes. Our current work mainly focus on a unified understanding to the various band edge changes of 2D BP.

TABLE SI. Orbital-pair projected COHP (-pCOHP) for interlayer P2-P3 interactions. The bold numbers represent the main orbital-pair contributions to interlayer interactions of CB@Γ, while that in Table II in main text summarizes all the numbers in this Table.

| | -*p*COHP | P3-3s | P3-3p$_y$ | P3-3p$_z$ | P3-3p$_x$ |
|---|---|---|---|---|---|
| CB@Γ | P2-3s | 0.000 | 0.003 | **-0.026** | 0.000 |
| | P2-3p$_y$ | 0.004 | -0.003 | **-0.009** | 0.000 |
| | P2-3p$_z$ | **-0.025** | **-0.010** | **0.041** | 0.001 |
| | P2-3p$_x$ | 0.000 | 0.001 | 0.000 | 0.000 |
| VB@Γ | P2-3s | 0.000 | -0.001 | -0.021 | 0.000 |
| | P2-3p$_y$ | -0.002 | 0.001 | -0.008 | 0.000 |
| | P2-3p$_z$ | -0.021 | -0.009 | -0.073 | 0.000 |
| | P2-3p$_x$ | 0.000 | 0.000 | 0.000 | 0.000 |

TABLE SII. Orbital-pair projected COHP (-pCOHP) for intralayer P2-P1 interactions. The bold numbers represent the main orbital-pair contributions to intralayer interactions of VB@Γ, while that in Table II in main text summarizes all the numbers in this Table.

| | -*p*COHP | P2-3s | P2-3p$_y$ | P2-3p$_z$ | P2-3p$_x$ |
|---|---|---|---|---|---|
| CB@Γ | P1-3s | -0.045 | -0.019 | -0.100 | 0.000 |
| | P1-3p$_y$ | -0.016 | -0.002 | -0.040 | 0.000 |
| | P1-3p$_z$ | -0.161 | -0.079 | -0.152 | 0.000 |
| | P1-3p$_x$ | 0.000 | 0.000 | 0.000 | -0.001 |
| VB@Γ | P1-3s | 0.012 | 0.006 | **-0.060** | 0.000 |
| | P1-3p$_y$ | 0.010 | 0.002 | **-0.053** | 0.000 |
| | P1-3p$_z$ | **-0.117** | **-0.061** | **0.246** | 0.000 |
| | P1-3p$_x$ | 0.000 | 0.000 | 0.000 | 0.000 |

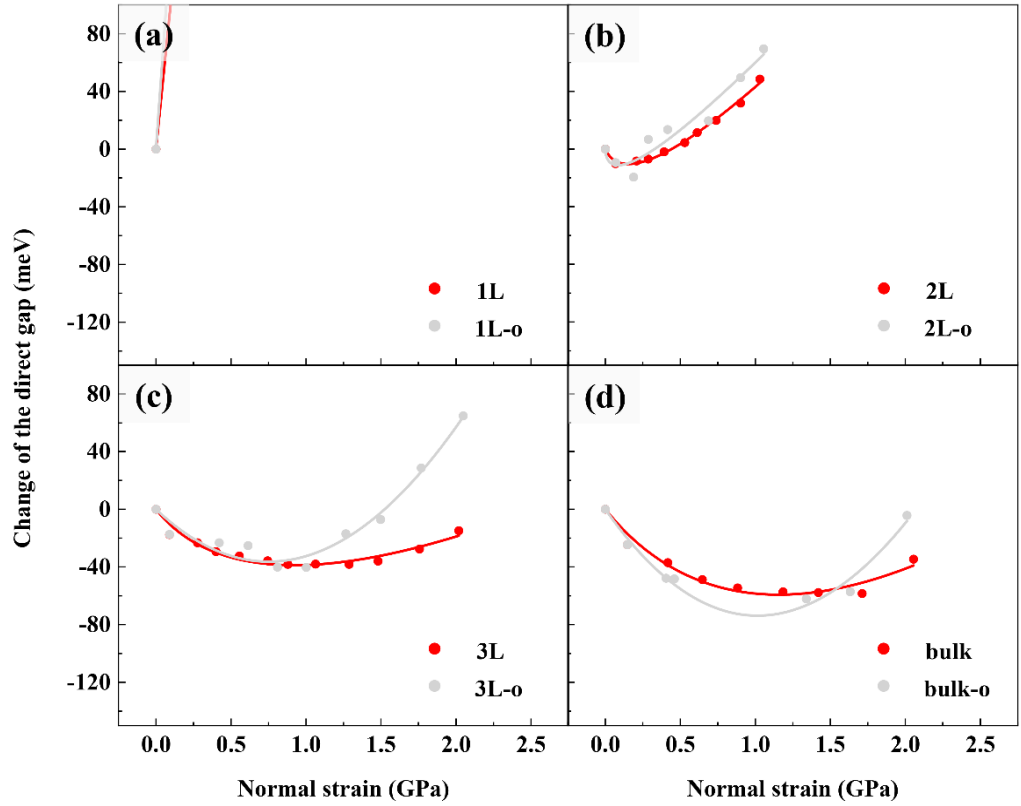

Fig. S1. Fixed vs. optimized (-o) in-plane lattice constant under normal strain.

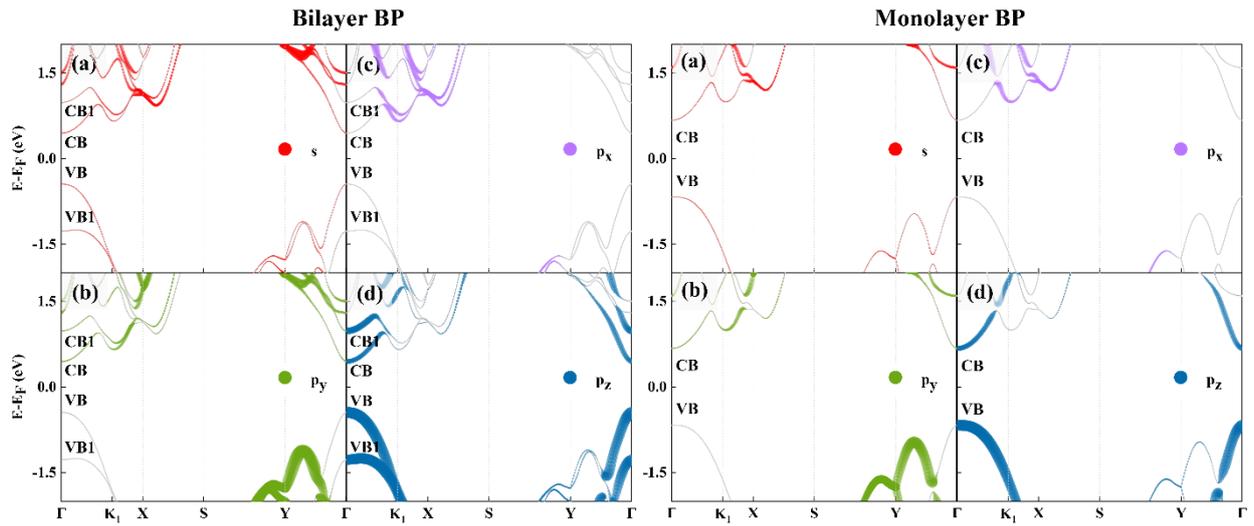

Fig. S2. Band structure projected to atomic-orbitals for bilayer and monolayer BP (zero pressure).

## CB@Γ of Monolayer BP

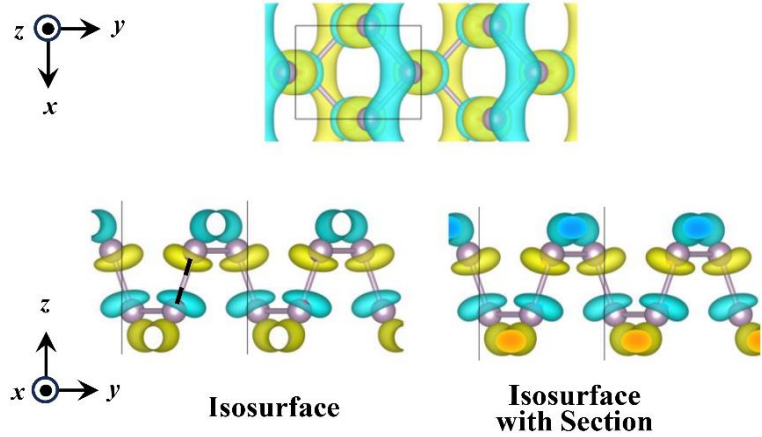

Fig. S3. Top- and side-views of the wave function real-space distribution of CB@Γ of monolayer BP. The side-views give isosurface without and with section.

## VB@Γ of Monolayer BP

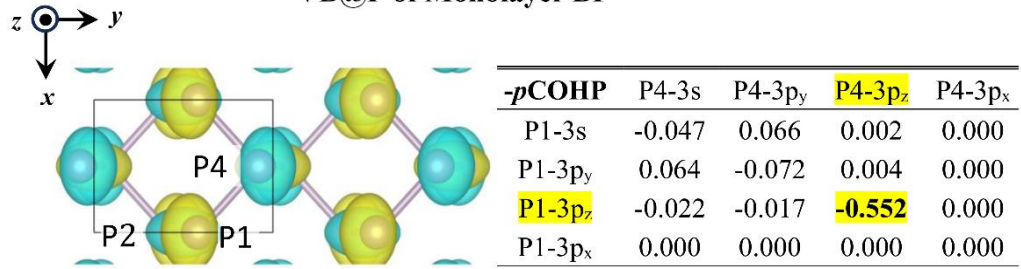

| -pCOHP | P4-3s | P4-3p$_y$ | P4-3p$_z$ | P4-3p$_x$ |
|---|---|---|---|---|
| P1-3s | -0.047 | 0.066 | 0.002 | 0.000 |
| P1-3p$_y$ | 0.064 | -0.072 | 0.004 | 0.000 |
| P1-3p$_z$ | -0.022 | -0.017 | **-0.552** | 0.000 |
| P1-3p$_x$ | 0.000 | 0.000 | 0.000 | 0.000 |

P1−P4: Antibonding of $p_z$−$p_z$

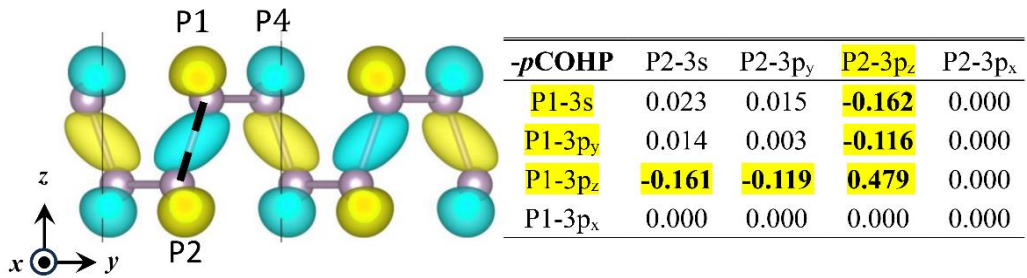

| -pCOHP | P2-3s | P2-3p$_y$ | P2-3p$_z$ | P2-3p$_x$ |
|---|---|---|---|---|
| P1-3s | 0.023 | 0.015 | **-0.162** | 0.000 |
| P1-3p$_y$ | 0.014 | 0.003 | **-0.116** | 0.000 |
| P1-3p$_z$ | **-0.161** | **-0.119** | **0.479** | 0.000 |
| P1-3p$_x$ | 0.000 | 0.000 | 0.000 | 0.000 |

P2-P1: Antibonding of $(s, p_y)$−$p_z$ & bonding of $p_z$−$p_z$; overall weak antibonding.

Fig. S4. Top- and side-views of the wave function of VB@Γ of monolayer BP along with the pCOHP analysis. The sum of P2-P1 -pCOHP values (-COHP = -0.024) in side-view indicates overall weak antibonding character between P2-P1.

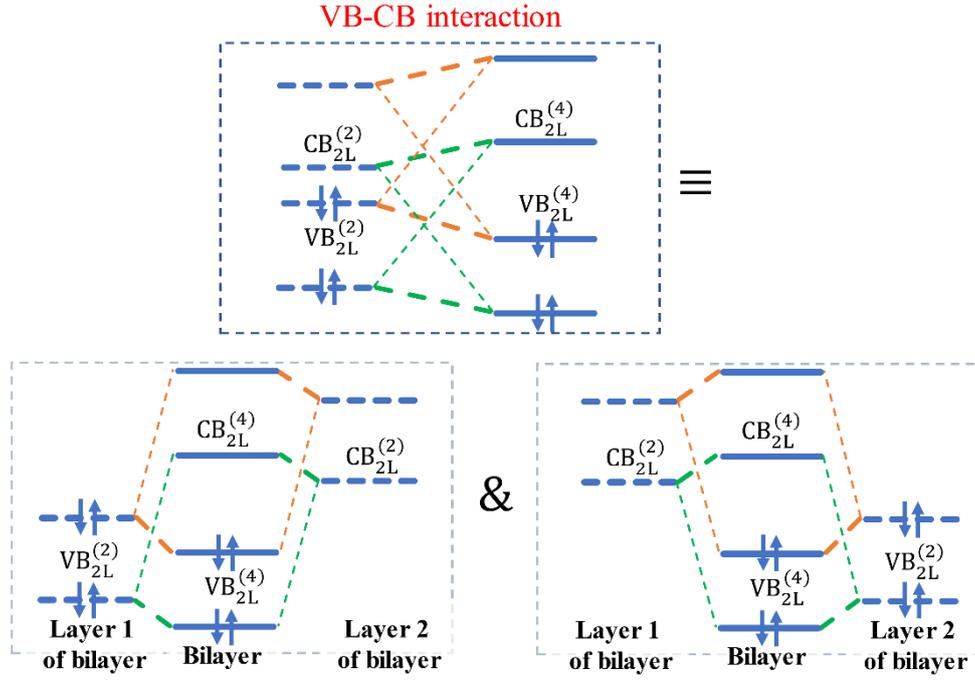

Fig. S5. The exact meaning of the occupied-empty interaction, in which $VB_{2L}^{(2)}$ and $CB_{2L}^{(2)}$ denote the band edges from the (imagined) two-level interactions, and, $VB_{2L}^{(4)}$ and $CB_{2L}^{(4)}$ indicate the final band edges from the four-level interaction. The simplified version (upper panel) is a resonance of the two lower-panels.

**CB@K$_1$ of Bilayer BP**

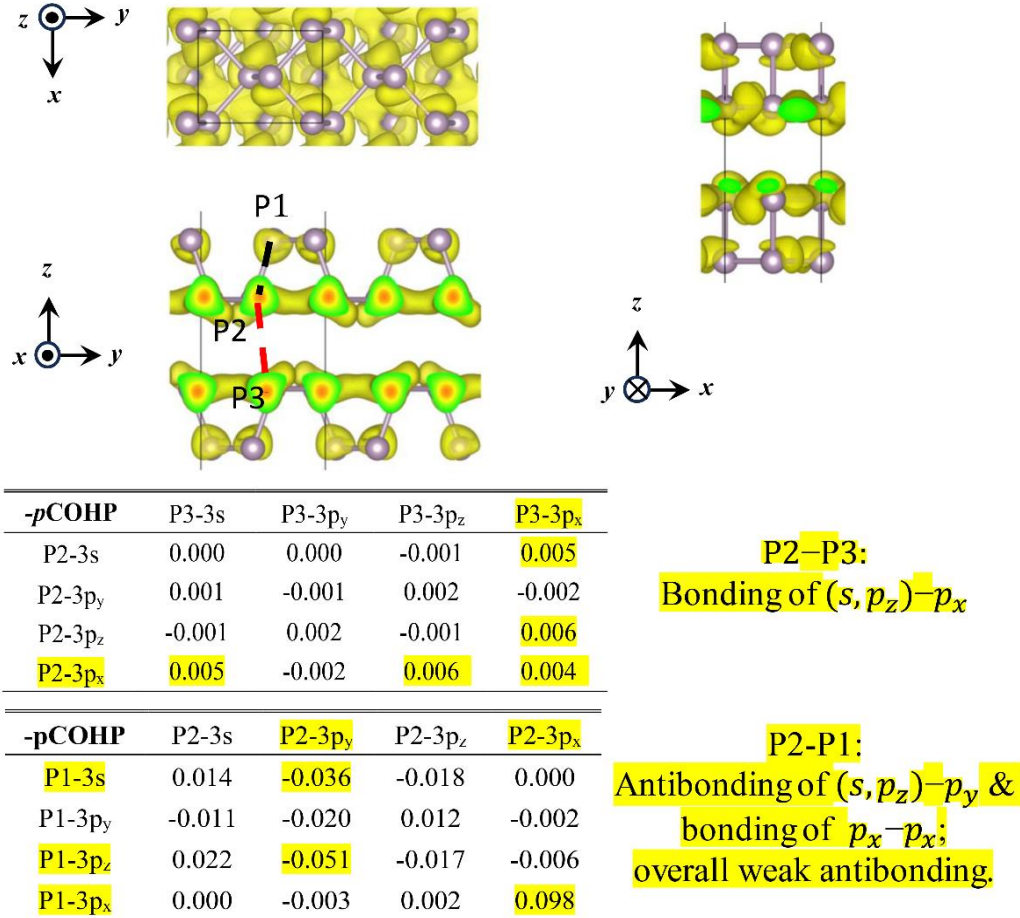

| -$p$COHP | P3-3s | P3-3p$_y$ | P3-3p$_z$ | P3-3p$_x$ |
|---|---|---|---|---|
| P2-3s | 0.000 | 0.000 | -0.001 | 0.005 |
| P2-3p$_y$ | 0.001 | -0.001 | 0.002 | -0.002 |
| P2-3p$_z$ | -0.001 | 0.002 | -0.001 | 0.006 |
| P2-3p$_x$ | 0.005 | -0.002 | 0.006 | 0.004 |

P2−P3:
Bonding of $(s, p_z)-p_x$

| -$p$COHP | P2-3s | P2-3p$_y$ | P2-3p$_z$ | P2-3p$_x$ |
|---|---|---|---|---|
| P1-3s | 0.014 | -0.036 | -0.018 | 0.000 |
| P1-3p$_y$ | -0.011 | -0.020 | 0.012 | -0.002 |
| P1-3p$_z$ | 0.022 | -0.051 | -0.017 | -0.006 |
| P1-3p$_x$ | 0.000 | -0.003 | 0.002 | 0.098 |

P2-P1:
Antibonding of $(s, p_z)-p_y$ &
bonding of $p_x-p_x$;
overall weak antibonding.

Fig. S6. Square of wave function modulus for CB@K$_1$ of bilayer BP. The sum of P2-P1 -$p$COHP values (-COHP = -0.016) shows overall weak antibonding character of intralayer P2-P1 interaction; and interlayer P2-P3 interaction shows bonding character from the summation of -$p$COHP values.

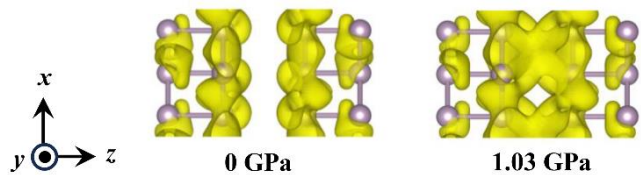

Fig. S7. Square of wave function modulus of the CB@K$_1$ state under normal strain for bilayer BP. Both figures are plotted with the same isosurface value.